\documentclass[twocolumn, aps]{revtex4-2}
\usepackage[latin9]{inputenc}
\usepackage{textcomp}
\usepackage{amsmath}
\usepackage{amssymb}
\usepackage{graphicx}
\usepackage{makecell}
\usepackage{float}
\usepackage{hyperref}
\usepackage{soul}
\makeatletter

\DeclareFontEncoding{LGR}{}{}

\ProvideTextCommand{\~}{LGR}[1]{\char126#1}

\@ifundefined{textcolor}{}{%
\definecolor{BLACK}{gray}{0}
\definecolor{WHITE}{gray}{1}
\definecolor{RED}{rgb}{1,0,0}
\definecolor{GREEN}{rgb}{0,1,0}
\definecolor{BLUE}{rgb}{0,0,1}
\definecolor{CYAN}{cmyk}{1,0,0,0}
\definecolor{MAGENTA}{cmyk}{0,1,0,0}
\definecolor{YELLOW}{cmyk}{0,0,1,0}
}

\usepackage{upgreek}
\usepackage{braket}
\usepackage{hyperref} 
\usepackage{xr-hyper}
\usepackage{physics}

\providecommand{\tabularnewline}{\\}

\@ifundefined{textcolor}{}{%
\definecolor{BLACK}{gray}{0}
\definecolor{WHITE}{gray}{1}
\definecolor{RED}{rgb}{1,0,0}
\definecolor{GREEN}{rgb}{0,1,0}
\definecolor{BLUE}{rgb}{0,0,1}
\definecolor{CYAN}{cmyk}{1,0,0,0}
\definecolor{MAGENTA}{cmyk}{0,1,0,0}
\definecolor{YELLOW}{cmyk}{0,0,1,0}
}

\usepackage{color}%\bibliographystyle{plain}
\usepackage{xcolor}
\def\NOT(#1,#2){\OneQubitGate(#1,#2){$X$}}

\@ifundefined{showcaptionsetup}{}{%
\PassOptionsToPackage{caption=false}{subfig}}

\makeatother

\begin{document}

\title{Nonlinear interferometry-based metrology of magneto-optical properties at infrared wavelengths}

\author{Tanmoy Chakraborty$^{1}$}

\author{Thomas Produit$^{1}$}

\author{Harish N S Krishnamoorthy$^{2,3}$}

\altaffiliation{Present address: Tata Institute of Fundamental Research Hyderabad, Gopanpally, Hyderabad 500046, Telangana, India}

\author{Cesare Soci$^{2,3}$}

\author{Anna V. Paterova$^{1}$}

\affiliation{$^{1}$A*STAR Quantum Innovation Centre (Q.InC), Institute of Materials Research and Engineering (IMRE), Agency for Science, Technology and Research (A$^{*}$STAR), 2 Fusionopolis Way, Innovis \#08-03, Singapore 138634, Republic of Singapore}

\affiliation{$^{2}$Centre for Disruptive Photonic Technologies, The Photonics Institute, Nanyang Technological University, 21 Nanyang Link, Singapore 637371, Republic of Singapore}

\affiliation{$^{3}$Division of Physics and Applied Physics, School of Physical and Mathematical Sciences, Nanyang Technological University, 21 Nanyang Link, Singapore 637371, Republic of Singapore}

\begin{abstract}

Magneto-optical properties of materials  are utilized in numerous applications both in scientific research and industries. The novel properties of these materials can be further investigated by performing metrology in the infrared wavelength range, thereby enriching their potential applications. However, current infrared metrology techniques can be challenging and resource-intensive due to the unavailability of suitable components. To address these challenges, we propose and demonstrate a set of measurements based on nonlinear interferometry, which allows us investigating magneto-optical properties of materials at infrared wavelength range by performing optical detection at the visible range. For a proof-of-principle study, we measure the Verdet constant of a bismuth-iron-garnet, over a spectral bandwidth of 600 nm in the near-IR range. 

\end{abstract}

\maketitle

%\textcolor{red}{check references}

\section{Introduction} 

Quantum metrology exploits the principles of quantum mechanics and allows measuring physical quantities with higher accuracy, better sensitivity and precision compared to conventional classical methods \cite{gobel2015quantum, barbieri2022optical, giovannetti2011advances, toth2014quantum, mukamel2020roadmap}. Research in quantum metrology has emerged substantially in recent years encompassing measurements of a broad variety of physical parameters. This includes extremely sensitive detection of fundamental constants of atoms and molecules \cite{rosenband2008frequency, morel2020determination}, probing magnetic and electric fields using ensembles of cold atoms \cite{vengalattore2007high, wildermuth2006sensing, behbood2013real, pezze2018quantum}, quantities associated with relativistic quantum field \cite{ahmadi2014quantum}, gravitational waves \cite{schnabel2010quantum, tse2019quantum, danilishin2019advanced} and dark matter \cite{backes2021quantum, dixit2021searching}. The present work is related to using quantum optics for metrology applications at infrared (IR) wavelength range, where a number of interesting physical properties of solid state and biological systems can be captured \cite{colarusso1998infrared, bhargava2012infrared}. Several applications involve Faraday rotation spectroscopy in the infrared part of the spectrum as well \cite{lewicki2009ultrasensitive, liu2023faraday}.

Conventional IR metrology techniques require light sources and detectors operating at IR wavelengths \cite{smith2011fundamentals, fard2013optical}. Hence, the performance of these techniques is limited by the significant inefficiency of existing IR detectors and the need for their operation at cryogenic temperatures, which reduces their accessibility and ease of use. The efficiency of the mercury cadmium telluride detectors is limited due to their high dark counts, low sensitivity even at cryogenic temperatures, and susceptibility to surrounding temperature fluctuations \cite{rogalski2012progress, lei2015progress}. Alternatively, IR spectroscopy can be performed by up-conversion of the IR signal, where an additional high power pump laser is employed to convert the wavelength of the IR photons to visible (or near-IR), where efficient detectors are easily available \cite{junaid2019video}. However, the detection performance in such systems is strongly constrained due to low conversion efficiency of nonlinear crystals \cite{israelsen2019real}.

Another method relies on the nonlinear interferometry of correlated photon pairs generated via spontaneous parametric down-conversion (SPDC) \cite{kalashnikov2016infrared} \cite{paterova2018measurement},  \cite{wang1991induced}, where one of the photons is generated at visible wavelength (signal), and the other at infrared (idler). Due to correlations between signal and idler SPDC photons, interference for signal photons is sensitive to the amplitude, phase and polarization of the electric field of idler photons. Thus, light-matter interactions at IR wavelength can be studied by probing the sample with idler photons and analyzing the interference for the signal photons at visible wavelengths. The main advantage of this method is that it does not require a detector for IR, the detection can be performed at visible wavelength, where efficient Si-based detectors are available. Moreover, a wide spectral tunability of the photons can be achieved by changing the phase matching within the same nonlinear crystal. The notion of "induced coherence without induced emission" was introduced by Wang \emph{el. al.} \cite{wang1991induced}, and later applied for various applications, including spectroscopy \cite{kalashnikov2016infrared, paterova2020nonlinear, kuznetsov2020nonlinear, paterova2017nonlinear, kutas2021quantum, burlakov1998three, kulik2004two, klyshko1993ramsey}, imaging \cite{lemos2014quantum, pearce2023practical, paterova2020quantum, paterova2020hyperspectral, yang2023interaction, santos2022subdiffraction}, as well as optical coherence tomography \cite{paterova2018tunable, valles2018optical, vanselow2020frequency,machado2020optical}. Polarization degree-of-freedom of the photons in such scheme is exploited for metrology applications \cite{grayson1994observation, paterova2019polarization}, tomography \cite{fuenzalida2024quantum} and quantum erasers \cite{herzog1995complementarity, gemmell2023coupling}. 

Through the present work we demonstrate a polarimetry method for measuring the magneto-optical properties of materials at IR wavelengths. Our technique is based on nonlinear interferometry of non-degenerate SPDC photon pairs that are generated by optically pumping a nonlinear crystal, which is embedded into the Michelson interferometer. The idler photon generated at IR wavelength serves as the probing photon, while the signal photon generated at a visible/near-IR wavelength is detected. In such a system, any change in the polarization of the idler photons introduced by the magneto-optical properties of the sample at IR wavelength is inferred by studying the interference of the signal photons. For a proof-of-principle demonstration of our method, we measure the Vedet constant and saturation Faraday rotation of a Bi$_3$Fe$_5$O$_{12}$ (BIG) crystal in the wavelength range of 1540 - 2141 nm by performing detection at 820 - 708 nm, respectively.

\section{Theoretical models}

Interference pattern of signal photons $I_{s}$ in a nonlinear interferometer with optically active media at the idler arm of the interferometer is expressed as follows \cite{paterova2019polarization} :
\begin{equation}\label{eq:vis_with_tau}
I_{\mathrm{s}} \propto 1+{|\tau_{i}|^{2}}|t||\mu|\cos(\phi_{p}-\phi_{i}-\phi_{s}+\Phi),
\end{equation}

where $\abs{\tau_{i}}$ is amplitude transmission coefficient for the idler photons through the optical components ($\abs{\tau_{i}}^{2}$ accounts for double pass of infrared photons though the specimen), $|t|$ determines polarization transfer, $\abs{\mu}$ is the normalized correlation function for the SPDC photons, $\phi_{p}$, $\phi_{s}$, $\phi_{i}$ are acquired phases by pump, signal and idler photons, respectively \cite{paterova2018measurement}, $\Phi$ is the initial phase in the interferometer.   

The visibility $\mathcal{V}$ of the interference pattern is defined from the maxima and minima of interference fringes and is proportional to the losses in the interferometer $|\tau_{i}|^{2}$, polarization transfer $|t|$ and correlation function $\abs{\mu}$:
\begin{equation}\label{eq:vis_def}
\mathcal{V} = \frac{{I_{\mathrm{s}}^{\mathrm{max}} - I_{\mathrm{s}}^{\mathrm{min}}}}{{I_{\mathrm{s}}^{\mathrm{max}} + I_{\mathrm{s}}^{\mathrm{min}}}} \propto {|\tau_{i}|^{2}}|t||\mu|.
\end{equation}

\begin{figure}
\includegraphics[width=1\columnwidth]{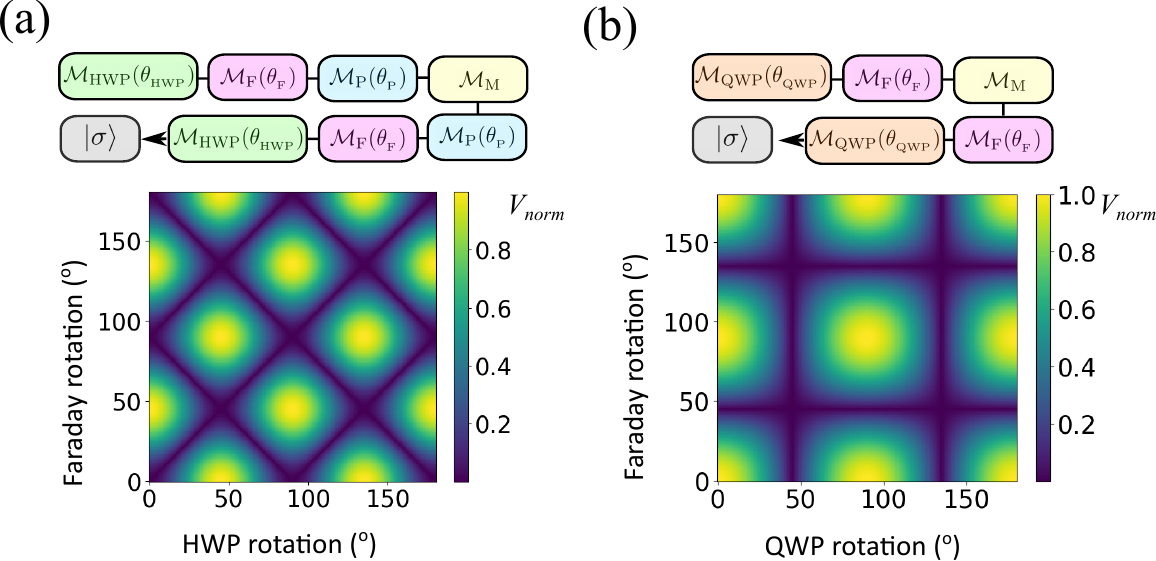}
\caption{Graphical representation of Jones matrix formalism and simulated 2-dimensional mapping showing variation of normalized visibility as a function of Faraday rotation angle introduced by the sample and the rotation angle of the (a) half-waveplate (with polarizer position fixed) and (b) quarter-waveplate. 
\label{fig:simulations}}
\end{figure}

The parameter $|t|$ is calculated employing Jones matrix formalism \cite{paterova2019polarization}. Here we use combinations of half- (HWP), quarter- (QWP) waveplates and a polarizer to observe the Faraday rotation introduced by a sample and construct two schemes for the polarimetry measurements. For case I, the idler photons  propagate through the HWP, the sample, the polarizer, then reflect from the mirror and travel back along the same path [see Fig. \ref{fig:simulations}(a)]. For the case II, the idler photons travel through the QWP and the sample, before reflecting back from the mirror [see Fig. \ref{fig:simulations}(b)].  Thus, the parameter $|t|$ is defined by transformation matrices $\mathcal{M}_{\mathrm{HWP}}({\theta}_{_{\mathrm{HWP}}})$, $\mathcal{M}_{\mathrm{QWP}}({\theta}_{_{\mathrm{QWP}}})$, $\mathcal{M}_{\mathrm{P}}({\theta}_{_{\mathrm{P}}})$, $\mathcal{M}_{\mathrm{F}}({\theta}_{_{\mathrm{F}}})$ and $\mathcal{M}_{\mathrm{M}}$ for HWP, QWP, polarizer, the sample with Faraday rotation and mirror, respectively (see Appendix A). Here $\theta_{_{\mathrm{HWP}}}$, $\theta_{_{\mathrm{QWP}}}$, $\theta_{_{\mathrm{F}}}$, and ${\theta}_{_{\mathrm{P}}}$ are the polarization rotation angles of the idler photons introduced by HWP, QWP, the sample and the polarizer, respectively.  Thus, in the scheme for case I the combined Jones matrix after double pass through the optical components is given by:
\begin{equation}
J_{_{\mathrm{I}}}=\mathcal{M}_{\mathrm{HWP}}\, \mathcal{M}_{\mathrm{F}}\,\mathcal{M}_{\mathrm{P}}\,\mathcal{M}_{\mathrm{M}}\,\mathcal{M}_{\mathrm{P}}\,\mathcal{M}_{\mathrm{F}}\,\mathcal{M}_{\mathrm{HWP}}.
\end{equation}

The initial state $|\sigma \rangle$ of vertically polarized idler photons can be written as $|\sigma \rangle=\begin{pmatrix}0 & 1\end{pmatrix}^\intercal$. Hence, the resulting polarization state of the idler photons is described as:
\begin{equation} \label{}
|\sigma\rangle_{_{\mathrm{I}}}=J_{_{\mathrm{I}}}|\sigma\rangle
\end{equation}

Since interference visibility for the signal photons is proportional to $|t|$ from Eq. \ref{eq:vis_with_tau}, the visibility for the interference of signal photons in this scheme is defined as a function of both $\theta_{_{\mathrm{HWP}}}$ and $\theta_{\mathrm{F}}$, see Appendix A for more details:
\begin{equation}\label{eq:Vis_HWP}
\mathcal{V}^{\mathrm{I}}=|\cos(2\theta_{_{\mathrm{HWP}}}+\theta_{\mathrm{F}}) \cos(2\theta_{_{\mathrm{HWP}}}-\theta_{\mathrm{F}})|.
\end{equation}

Next, the combined Jones matrix for case II can be written as:
\begin{equation}
J_{_{\mathrm{II}}}=\mathcal{M_{\mathrm{QWP}}}\,\mathcal{M}_{\mathrm{F}}\,\mathcal{M}_{\mathrm{M}}\,\mathcal{M}_{\mathrm{F}}\,\mathcal{M}_{\mathrm{QWP}}.
\end{equation}

Then, in this case the normalized visibility as a function of $\theta_{_{\mathrm{QWP}}}$ and $\theta_{\mathcal{\mathrm{F}}}$ is given by:
\begin{equation}\label{eq:Vis_QWP}
\mathcal{V}^{\mathrm{II}}=|\cos(\theta_{_{\mathrm{QWP}}})\cos(\theta_{\mathrm{F}})|.
\end{equation}

Figure \ref{fig:simulations}(a) [(b)] shows the mapping of normalized interference visibility as a function of the Faraday rotation angle by the sample and the angle of rotation of the HWP [QWP], calculated  using Eq.(\ref{eq:Vis_HWP}) [Eq.(\ref{eq:Vis_QWP})]. In both visibility maps, the higher values correspond to better indistinguishability of the idler photons, which is discussed more in details in the following sections. In the experiments the visibilities can be measured depending on the orientation of the waveplates for a given applied magnetic field (or the introduced Faraday rotation). These measurements correspond to the horizontal cross-sections of the graphs in Fig. \ref{fig:simulations}.

\section{Experimental setup and measurement protocols}

\begin{figure}
\includegraphics[width=0.9\columnwidth]{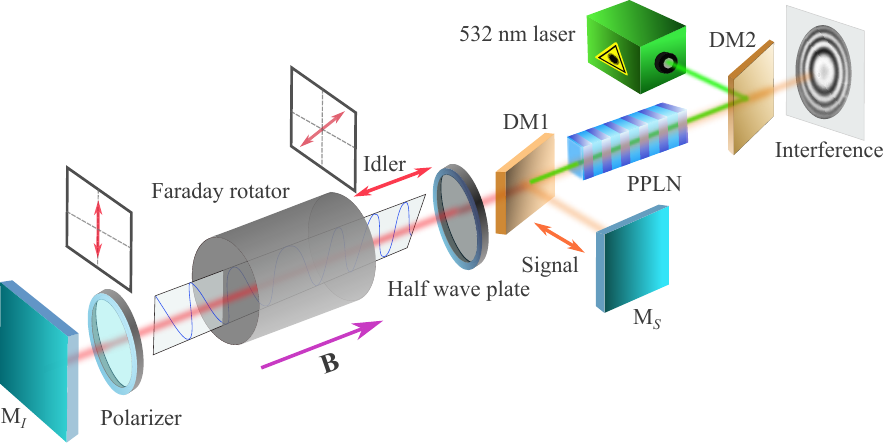}
\caption{A simplified schematic of the experimental setup based on Michelson interferometer. PPLN: periodically poled lithium niobate; DM1: dichroic mirror that transmits idler photons and reflects signal and pump photons; DM2: dichroic mirror that reflects pump and transmits signal; M$_s$ and M$_i$: signal and idler mirrors, respectively. Interference of the signal photons are captured using the sCMOS camera. 
\label{fig:setup}}
\end{figure}

A simplified schematic of our experimental setup is shown in Fig. \ref{fig:setup}. Signal-idler SPDC photon pairs are generated by focusing the continuous wave green laser (532 nm wavelength, 40 mW power, Laser Quantum) pump beam into a periodically poled lithium niobate (PPLN) nonlinear crystal (Covesion). The PPLN crystal consists of multiple periodic poling channels, each designed to meet different quasi-phase matching (QPM) conditions. QPM via different poling and temperature tuning allows generating idler photons over 1500 nm - 2300 nm wavelength range, while conjugate signal photons are generated at 824 nm to 692 nm. After the PPLN crystal the co-propagating signal and idler beams are separated using a dichroic mirror (DM1) and collimated using two lenses, each with $f$=125 mm focal length. The signal and idler photons propagate along the two arms of the Michelson interferometer and reflect back from two metallic mirrors. The mirror M$_i$ at the idler arm is mounted on a motorized translation stage, which allows to introduce the additional phase for the idler photons. The pump beam following the signal path is reflected back by the mirror M$_s$ into the PPLN crystal, which generates the signal and idler SPDC photons for the second time. After separating the signal and pump photons by a second dichroic mirror (DM2), the signal photons are focused onto a sCMOS camera (Thorlabs, Inc.). At the input of the sCMOS camera we use a notch at 532 nm and a band-pass filters to block any residual pump photons. 

The BIG sample under study is placed at the path of idler photons (see the crystal structure of the sample and its magnetization properties given in Appendix B). An externally applied magnetic field $B$ to the sample along the optical axis results in Faraday rotation of the polarization of idler photons. Below the saturation field, the Faraday rotation angle ${\Theta}_{\mathrm{F}}$ increases linearly with the strength of the applied magnetic field as follows:
\begin{equation}
\label{eq:Verdet_constant}
{\Theta}_{\mathrm{F}}=V(\lambda)Bl,
\end{equation}

where $l$ is length of the crystal along the light propagation direction (in our case $l = $0.38 mm). Thus, at the linear regime of the Faraday rotation it is possible to measure the Verdet constant $V(\lambda)$.

To detect the Faraday rotation introduced by the sample we use two different configurations of the idler arm of the interferometer. In case I, after the DM1 idler photons pass through the HWP, BIG sample and polarizer. The HWP introduces a known polarization transformation to idler photons, which is followed by Faraday rotation from the BIG crystal. Subsequently, the polarizer retrieves the polarization back to its initial state, as it was before the HWP. After reflecting back from M$_i$, the idler photons travel back along the same path through the polarizer, BIG and HWP. The polarization state is not changed by the polarizer, though it is altered by the Faraday rotation in the BIG crystal and the retardation introduced by the HWP. In case II, a known amount of retardation is introduced to the idler photons by QWP, which follows the Faraday rotation by the BIG crystal. These two operations are then applied in reverse order after the photons are reflected back by mirror M$_i$. 

We analyze the resulting polarization state of the idler photons by measuring the visibilities of the interference fringes for the signal photons. These fringes are observed via introducing a relative phase by translating the mirror M$_i$ along the optical axis of the idler arm. Thus, by capturing the variation of visibility pattern as a function of Faraday rotation angles at different applied magnetic fields $B$, we study the magneto-optical properties of BIG at IR wavelengths, according to Eqs. \ref{eq:Vis_HWP}, \ref{eq:Vis_QWP} and \ref{eq:Verdet_constant}.

\section{Experimental results}

First, we pump the PPLN crystal along a polling period that allows generating the signal and idler photons at 813 nm and 1540 nm, respectively (see the signal spectrum in Appendix C). The full width at half maximum of the spectrum is 1.4$\pm$0.02 nm, which gives us $l_{_{\mathrm{coh}}}=0.48\pm0.01$ mm coherence length. We measure the intensity of signal photons across $\approx$ 0.8 mm path length difference without and with the sample at the path of idler photons (see Figs. \ref{fig:Coh_length_no_BIG} and \ref{fig:Coh_length_with_BIG} in Appendix C) and confirm the calculated coherence length experimentally. The inclusion of the BIG crystal into the path of idler arm changes the optical path length to $\approx$ 0.53 mm, which depends on the refractive index of BIG and its thickness $l$. In this case the optical  path length difference can be compensated by shifting the position of the mirror M$_i$. 

First, we perform interference measurements for case I and II without introducing the BIG crystal into the interferometer (see  Figs. \ref{fig:V_without_BIG _HWP} and \ref{fig:V_without_BIG_QWP} in Appendix D for the case I and II, respectively). The visibility of the interference fringes depends on the orientation of the waveplates (${\theta}_{\mathrm{HWP}}$ for case I and ${\theta}_{\mathrm{QWP}}$ for case II) as calculated by our theoretical model in Eq. \ref{eq:Vis_HWP} for case I and Eq. \ref{eq:Vis_HWP} for case II. Next, we introduce the BIG sample, balance the interferometer arms and perform the polarimetry measurements.

\subsection{Results for case I: with HWP and polarizer}

\begin{figure}[h]
\includegraphics[width=0.75\columnwidth]{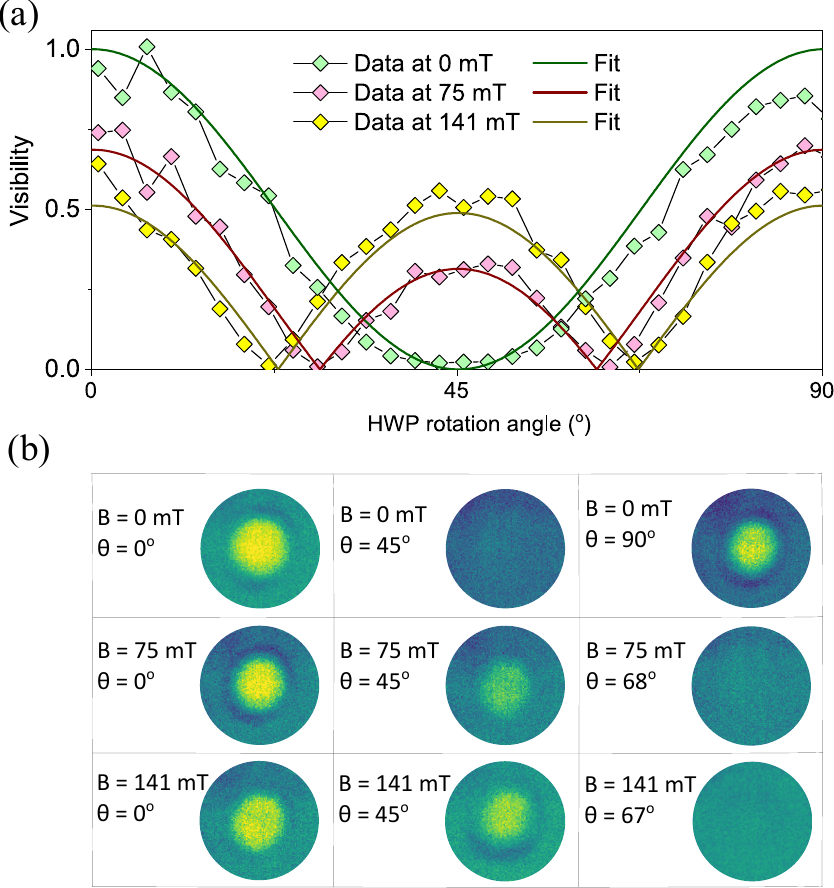}
\caption{(a) Visibility of interference fringes of signal photons depending on the rotation angle of the HWP at different magnetic field strengths applied to the BIG crystal. (b) Spatial intensity distribution of the signal photons revealing constructive interference patterns captured using the sCMOS camera. A few of such patterns are shown for different magnetic field strengths and HWP rotation angles.
\label{fig:HWPdata}}
\end{figure}

We measure the visibility of the interference fringes for the signal photons with the orientation of the HWP at zero magnetic field $B=0$ mT, which is shown in Fig. \ref{fig:HWPdata}(a). The normalized visibility $\mathcal{V}$ is maximum when ${\theta}_{\mathrm{HWP}}=0^{\circ}$. It reaches minimum value when ${\theta}_{\mathrm{HWP}}=45^{\circ}$, and restores its maximum  again when ${\theta}_{\mathrm{HWP}}=90^{\circ}$, which is in accordance with Eq. \ref{eq:Vis_HWP} for ${\theta}_{\mathrm{F}}=0^{\circ}$.

The visibility as a function of ${\theta}_{\mathrm{HWP}}$ for different strengths of the magnetic field is shown in Fig. \ref{fig:HWPdata}(a) as well. Interestingly, initially one minimum in visibility curve at $B=0$ mT splits into two for $B\neq0$ mT. The separation of these minima increases with the strength of the applied magnetic field and reaches maximum for the saturation field. These results are consistent with the theoretical model given by Eq. \ref{eq:Vis_HWP}. We fit the experimental data for $B=75$ mT and 141 mT with keeping ${\theta}_{\mathrm{F}}$ as the fitting parameter, which yields ${\theta}_F=34.08^{\circ}\pm0.22^{\circ}$ and  $44.34^{\circ}\pm0.46^{\circ}$, respectively. 

Change in visibility of the interference pattern is associated with indistinguishability of SPDC photons generated in two passes of the pump beam through the nonlinear crystal. The waveplate and Faraday rotator change the polarization state of idler photons, which reduces the degree of indistinguishability and hence, the interference visibility of signal photons \cite{paterova2019polarization}. The stronger magnetic field causes larger Faraday rotation, reducing the degree of indistinguishability further. As an example, Fig. \ref{fig:HWPdata}(b) shows a few measured spatial intensity distributions of signal photons for different values of ${\theta}_{\mathrm{HWP}}$ and $B$. 

\subsection{Results for case II: with QWP}

\begin{figure}[h]
\includegraphics[width=0.75\columnwidth]{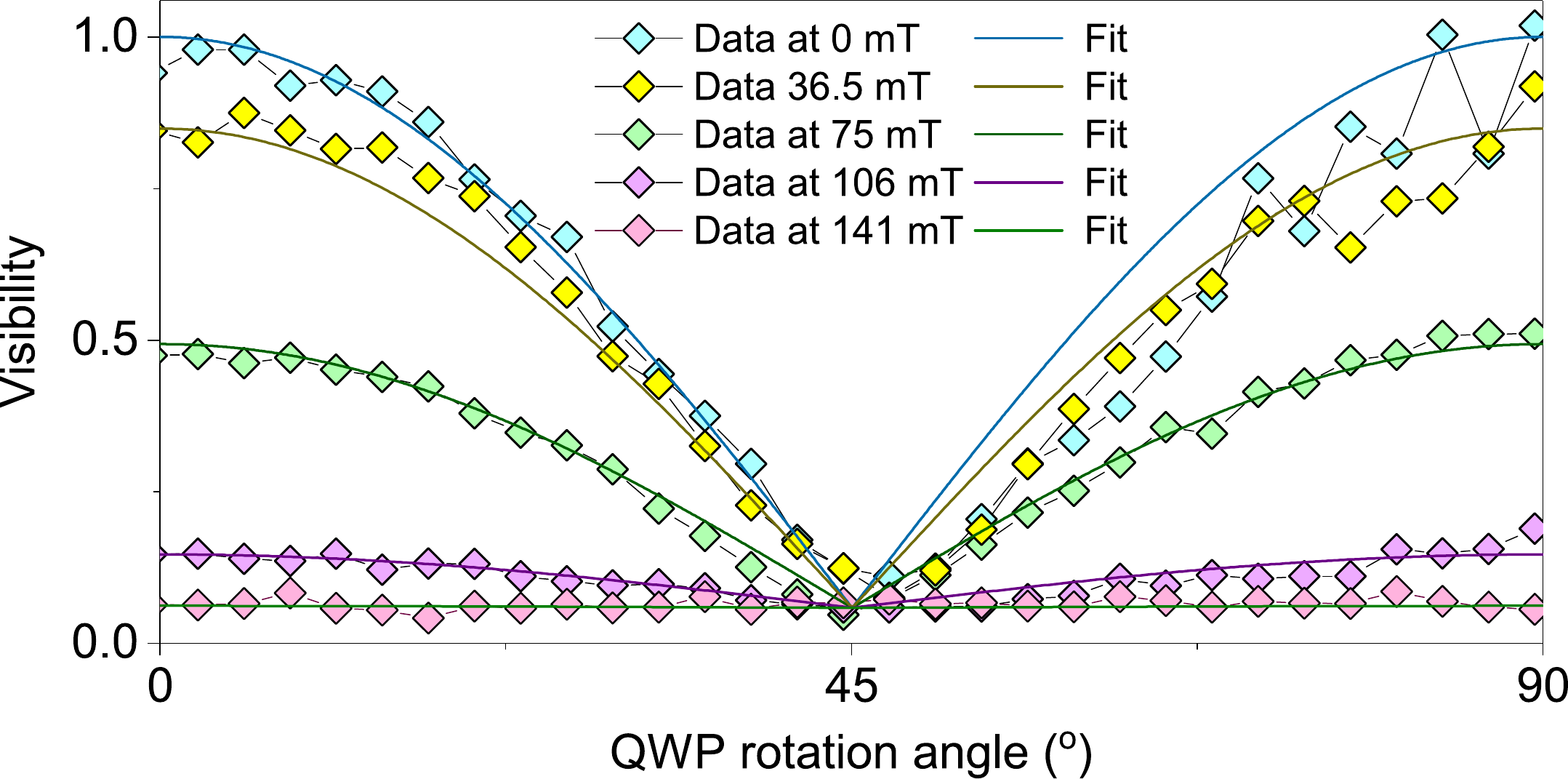}
\caption{Visibility of the interference fringes for the signal photons measured as a function of the rotation angle of QWP for different magnetic field strengths, causing different amount of Faraday rotations. 
\label{fig:QWPdata}}
\end{figure}

In this scheme, we replace the HWP shown in Fig. \ref{fig:setup} with a QWP, remove the polarizer and perform manipulation of the input polarization of idler photons by rotating the QWP. The dependence of the visibility of the interference fringes on the orientation of the QWP for different values of $B$ is shown in Fig. \ref{fig:QWPdata}. For $B=0$ mT and $\theta_{\mathrm{QWP}}=0^{\circ}$, the photons generated during the forward and backward pass of the pump are indistinguishable, being in the same polarization state. However, the degree of indistinguishability is reduced with increasing $\theta_{QWP}$, and the visibility of  the interference fringes reaches minimum when $\theta_{QWP}=45^{\circ}$. Additional polarization rotation introduced by the BIG varies with the strength of the magnetic field $B$, and the maximum visibility of the interference fringes are reduced with increasing the magnetic field strength. Visibility curves depending on the orientation of the QWP for each of the values of $B$ are fitted to  Eq. \ref{eq:Vis_QWP}, wheres, the Faraday rotation angles caused by the BIG crystal are determined (see the summarized Faraday rotation angles depending on the applied magnetic field in table \ref{table:QWP_BIG_rot} in Appendix E). For the applied magnetic fields $B=75$ mT and 141 mT the resulting Faraday rotation angles are ${\theta}_F=31.24^{\circ}\pm0.15^{\circ}$ and  $44.9^{\circ}\pm0.45^{\circ}$, respectively, which agree with the measured values in case I.

\subsection{Wavelength dependent Faraday rotation}

\begin{figure}[t!]
\includegraphics[width=0.55\columnwidth]{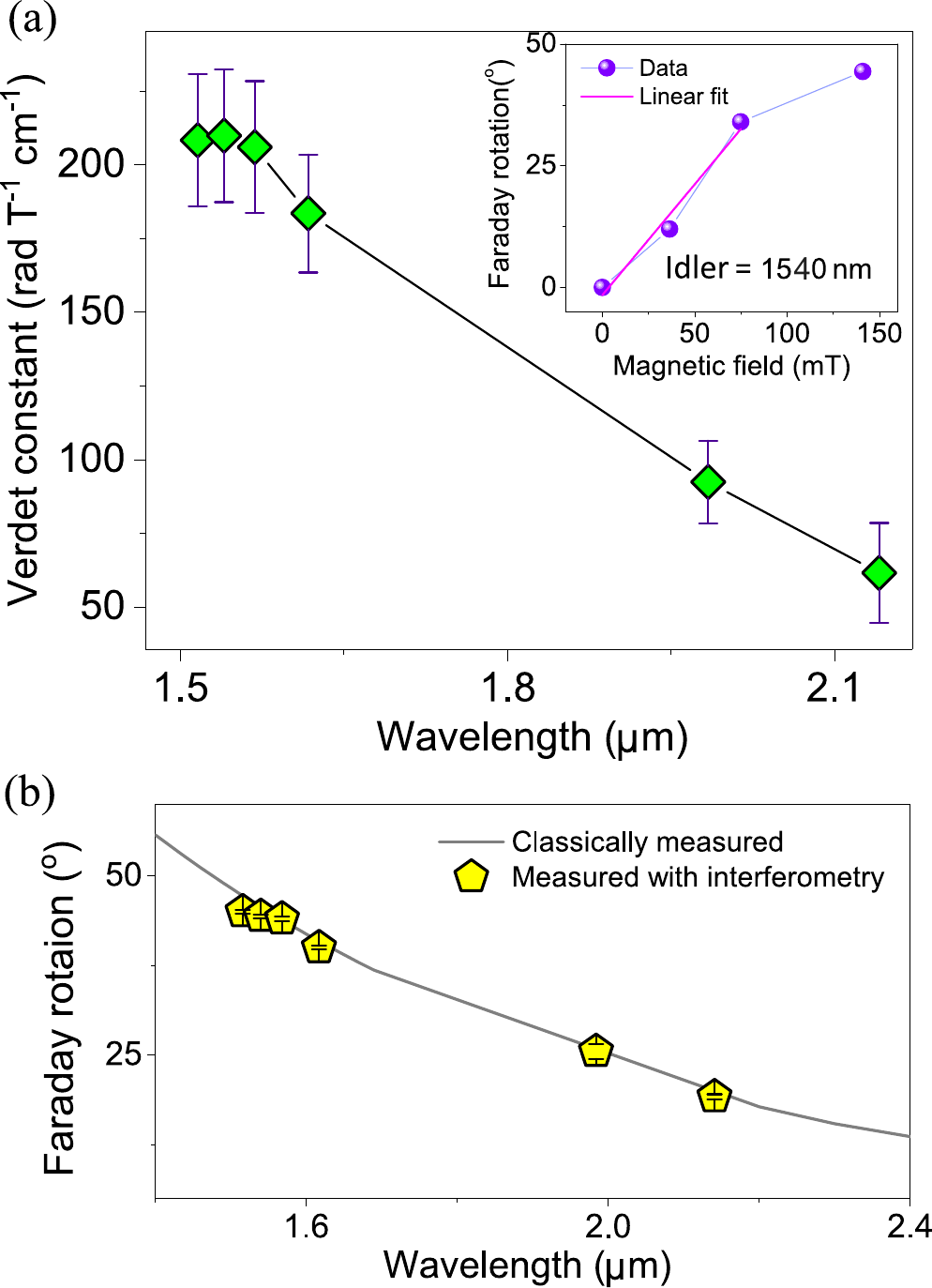}
\caption{(a) The Verdet constant depending on wavelength. Points correspond to data measured employing the interferometry scheme.  Inset shows Faraday rotations caused by the BIG crystal with applied magnetic field and its linear fit for $\lambda_{_{\mathrm{\mathrm{idler}}}}$=1540 nm. (b) Maximum Faraday rotation (at saturation magnetic field) introduced by the BIG sample depending on wavelength: the hexagons represent the experimental data measured in interferometry scheme, and the solid line show the values provided by manufacturer. 
\label{fig:Verdet_const}}
\end{figure}

 Next, we generate the idler photons at IR wavelength range from $\lambda_{\mathrm{idler}}$ = 1516 nm to 2141 nm. The coarse tuning of $\lambda_{\mathrm{idler}}$ is done by changing the poling period and fine-tuning is done by changing the applied temperature to the PPLN crystal while being in the same poling channel. Following the energy conservation law, the detected signal photons are generated in the range of 708 nm to 820 nm (see table 1 in Appendix F for the details). To capture the dependence of  the Verdet constant $V(\lambda)$ on wavelength we choose the scheme in case I of the interferometer, as both the maxima and the minima of the visibility undergoes changes with the applied magnetic field. For each wavelength we capture Faraday rotations for different strengths of the applied magnetic field $B$. Following same protocol as for the data captured at $\lambda$=1540 nm, we measure $V(\lambda)$ at IR wavelengths over 600 nm bandwidth range [see Fig. \ref{fig:Verdet_const}(a)]. 

The estimated Faraday rotation angles for case I are plotted depending on $B$ in the inset of Fig. \ref{fig:Verdet_const}. The data that are below the saturation field $B_{sat}$ is fitted as a linear curve. The slope of the fit allows estimating the Verdet constant $V(\lambda)$ for BIG crystal at $\lambda$=1540 nm. Note that beyond the saturation field at $B=150$ mT the Faraday rotation angle is ${\theta}_F=45.0^{\circ}\pm0.5^{\circ}$, which corresponds to the supposed specifications. The plot shows that $V(\lambda)$ decreases with the wavelength and becomes significantly small at 2141 nm. From the acquired data we also estimate the Faraday rotation angles at saturation magnetic field, which is shown the in Fig. \ref{fig:Verdet_const}(b). The data is well consistent with their classically measured counterparts.

\section{Discussion and conclusions}

To summarize, we have demonstrated a metrology scheme based on a nonlinear interferometry to investigate the magneto-optical properties of a specimen at IR wavelengths. The method is implemented by using a Michelson interferometer, where the paths of the signal and idler SPDC photons construct the two arms of the interferometer. Through our experiments, a polarization rotation of the idler photons introduced by the sample at IR wavelength is inferred by studying the interference of the signal photons. This allows performing measurement of magneto-optical properties of the sample at IR wavelengths by performing detection at visible or near-IR wavelengths. 

For a proof-of-concept demonstration of our method, we performed metrology of magneto-optical properties of Bi$_3$Fe$_5$O$_{12}$ crystal, which shows detectable Faraday rotation at IR wavelength range. We formulated theoretical analysis for calculating the visibility of the interference for signal photons as a function of the polarization rotation introduced by the sample at the idler photons wavelength. We captured the Verdet constant and Faraday rotation angles over a bandwidth from 1516 to 2141 nm, while the detection is performed at the range from 820 nm to 708 nm. The results obtained through our method show consistency with the classically measured data, which proves the applicability of our method. By choosing different QPM conditions in PPLN crystal used in the experiments, the metrology experiments can be performed at even longer wavelengths, up to 5 $\mu$m, which is given by the transparency range of the PPLN crystal. The method can be extended even further towards measuring materials which show detectable Faraday rotation at longer wavelengths (5-10 $\mu$m) using suitable nonlinear crystals like silver gallium sulfide \cite{kumar2021mid} and gallium phosphide \cite{wilson2020integrated}.

The proposed IR metrology protocol and its experimental demonstration in the present paper open up possibilities for unleashing magneto-optical properties of a large class of materials, that is not yet well explored due to technical challenges. Such investigations can reveal novel phenomena of these materials in the IR wavelength range, which can significantly boost their applications in scientific research and industries. IR polarimetry can have applications in defence and facial recognition  at night \cite{bieszczad2021review} and in characterization of painting materials \cite{hagen2022review}.
These rely on polarimetry of reflection and thermal emission from surfaces, which is rather complex to analyse. Such measurements can be benefited from our IR polarimetry technique. Our method also provides a new path towards investigating light-matter interaction phenomena in novel material systems such as topological insulators, which have exotic surface states that are sensitive to the helicity of light \cite{krishnamoorthy2023topological,spektor2015spin}.

\section{Acknowledgements}

We thank Jian Rui Soh and Vytautas Valuckas for their help with characterizing the sample.
AP, TP and TC acknowledge support to this research A*STAR  through the grant no. C230917004. and for support by NRF Singapore through the grant NRF2021-QEP2-03-P08.

\maketitle

\section*{Appendix A: Calculation of the interference visibilities for schemes of the interferometer in \emph{case I} and \emph{case II}}

To calculate the parameter $t$ expressed in Eq. \ref{eq:vis_with_tau}, we first determine the transfer matrix for the optical components described in section III for cases I and II. Change in the polarization of the idler photons by each of the optical components is determined by employing Jones matrix formalism. According to the schemes of the interferometer we express Jones matrices for half- and quarter- wave plates, Faraday rotator, polarizer, and a mirror as: $\mathcal{M}({\theta}_{_{\mathrm{HWP}}})$, $\mathcal{M}({\theta}_{_{\mathrm{QWP}}})$, $\mathcal{M}({\theta}_{_{\mathrm{F}}})$, $\mathcal{M}_{\mathrm{P}}$ and $\mathcal{M}_{\mathrm{M}}$, respectively:

\begin{equation} \label{eq:M_HWP}
\mathcal{M}_{\mathrm{HWP}}=
\begin{pmatrix}
\cos2\theta_{_{\mathrm{HWP}}} & \sin2\theta_{_{\mathrm{HWP}}} \\
\sin2\theta_{_{\mathrm{HWP}}} & -\cos2\theta_{_{\mathrm{HWP}}} 
\end{pmatrix} ,
\end{equation}

\begin{equation} \label{eq:M_QWP}
\mathcal{M}_{\mathrm{QWP}}= \frac{1}{\sqrt{2}}
\begin{pmatrix}
1 + i\cos2\theta_{_{\mathrm{QWP}}} & i\sin2 \theta_{_{\mathrm{QWP}}} \\
i\sin2\theta_{_{\mathrm{QWP}}} & 1 - i\cos2 \theta_{_{\mathrm{QWP}}}
\end{pmatrix},
\end{equation}

\begin{equation} \label{eq:M_F}
\mathcal{M}_{\mathrm{F}}=
\begin{pmatrix}
\cos\theta_{_{\mathrm{F}}} & \sin\theta_{_{\mathrm{F}}} \\
-\sin\theta_{_{\mathrm{F}}} &  \cos\theta_{_{\mathrm{F}}} 
\end{pmatrix} ,
\end{equation}

\begin{equation} \label{eq:M_P}
\mathcal{M}_{\mathrm{P}}=
\begin{pmatrix}
0 & 0 \\
0 &  1 
\end{pmatrix},   
\mathcal{M}_{\mathrm{M}}=
\begin{pmatrix}
1 & 0 \\
0 &  1 
\end{pmatrix},
\end{equation}

where angles $\theta_{_{\mathrm{HWP, QWP, F}}}$ denote angles between initial polarization state of the idler photons and an orientation of the waveplates, or Faraday rotation. In Eq. \ref{eq:M_P} the matrix for polarizer  $\mathcal{M}_{\mathrm{P}}$ is given for the vertical orientation, which is aligned with the initial polarization for the idler photons. The matrix $\mathcal{M}_{\mathrm{M}}$ is given by the unit matrix since the mirror does not alter the polarization of the propagating beams. 

In case I, the polarization is changed by the HWP, BIG sample and polarizer at the forward path along the arm of the Michelson interferometer: 
\begin{equation} \label{eq:M_F2}
\mathcal{M}_{_{1}}^{\mathrm{I}} = \mathcal{M}_{\mathrm{HWP}}\,\mathcal{M}_{\mathrm{F}}\,\mathcal{M}_{\mathrm{P}}=
\begin{pmatrix}
0 & \sin{(2\theta_{_{\mathrm{F}}} + \theta_{_{\mathrm{HWP}}})} \\
0 &  \cos{(2\theta_{_{\mathrm{F}}}+ \theta_{_{\mathrm{HWP}}})}
\end{pmatrix}. 
\end{equation}

After the reflection from M$_i$, the combined matrix is given in reversed order $\mathcal{M}_{_{2}}^{\mathrm{I}} = \mathcal{M}_{\mathrm{P}}\,\mathcal{M}_{\mathrm{F}}\,\mathcal{M}_{\mathrm{HWP}}$. Hence the total transfer matrix is defined as:
\begin{equation} \label{eq:M_F2}
\begin{aligned}
J_{\mathrm{I}}= \mathcal{M}_{_{1}}^{\mathrm{I}}\,\mathcal{M}_{\mathrm{M}}\,\mathcal{M}_{_{2}}^{\mathrm{I}} =   \\
\frac{1}{2}
\begin{pmatrix}
\cos{2\theta_{_{\mathrm{F}}}} 
- \cos{4\theta_{_{\mathrm{HWP}}}} 
& 
-\sin{2\theta_{_{\mathrm{F}}}} 
- \sin{4\theta_{_{\mathrm{HWP}}}} \\
\sin{2\theta_{_{\mathrm{F}}}} 
- \sin{4\theta_{_{\mathrm{HWP}}}}
&  
\cos{2\theta_{_{\mathrm{F}}}} 
+ \cos{4\theta_{_{\mathrm{HWP}}}}
\end{pmatrix}. 
\end{aligned}
\end{equation}

The initial state $|\sigma \rangle$ of vertically polarized idler photons can be written as $|\sigma \rangle=\begin{pmatrix}0 & 1\end{pmatrix}^\intercal$. The final state of the idler photons after using the Jones matrix and simplification of the equations is given by:
\begin{equation} \label{eq:M_F3}
|\sigma\rangle_{_{\mathrm{I}}} = J_{_{\mathrm{I}}} \begin{pmatrix} 0  \\ 1 \end{pmatrix}  = \frac{1}{2} \begin{pmatrix} -\sin{2\theta_{_{\mathrm{F}}}} 
- \sin{4\theta_{_{\mathrm{HWP}}}}  \\  \cos{2\theta_{_{\mathrm{F}}}} + \cos{4\theta_{_{\mathrm{HWP}}}}
\end{pmatrix}.
\end{equation}

The resulting interference depends on the vertical polarization component. Thus, the visibility of the interference fringes depends on the orientation of the HWP and the Faraday rotation as follows:
\begin{equation} \label{eq:M_F4}
\begin{aligned}
\mathcal{V}^{\mathrm{I}} = \frac{1}{2} |(\cos{2\theta_{_{\mathrm{F}}}} + \cos{4\theta_{_{\mathrm{HWP}}}})| = \\
|\cos{(2\theta_{_{\mathrm{HWP}}}
+ \theta_{_{\mathrm{F}}})} \cos{(2\theta_{_{\mathrm{HWP}}}
-\theta_{_{\mathrm{F}}})}|.
\end{aligned}
\end{equation} 

Similarly, in case II, the Jones matrices for combined QWP and BIG crystal towards M$_i$ and backwards from  M$_i$ are given respectively by: 
\begin{equation} \label{eq:M_F5}
\begin{aligned}
\mathcal{M}_{_{1}}^{\mathrm{II}} =\frac{\sqrt{2}}{2} \cdot  \\
\begin{pmatrix}
\cos{\theta_{_{\mathrm{F}}}} 
+ i\cos{(2\theta_{_{\mathrm{QWP}}} + \theta_{_{\mathrm{F}}})} 
& 
\sin{\theta_{_{\mathrm{F}}}} 
+ i\sin{(2\theta_{_{\mathrm{QWP}}} + \theta_{_{\mathrm{F}}})}  \\
-\sin{\theta_{_{\mathrm{F}}}} 
+ i\sin{(2\theta_{_{\mathrm{QWP}}} + \theta_{_{\mathrm{F}}})} 
&  
\cos{\theta_{_{\mathrm{F}}}} 
- i\cos{(2\theta_{_{\mathrm{QWP}}} + \theta_{_{\mathrm{F}}})} 
\end{pmatrix}. 
\end{aligned}
\end{equation}

Hence, the Jones matrix for double-pass configuration is  $J_{\mathrm{II}} = \mathcal{M}_{_{1}}^{\mathrm{II}}\,\mathcal{M}_{\mathrm{M}}\,\mathcal{M}_{_{2}}^{\mathrm{II}}$.
The final visibility depending on the orientation of the QWP and the Faraday rotation by the BIG sample is given by:
\begin{equation}\label{eq:Vis_QWP1}
\mathcal{V}^{\mathrm{II}}=|\cos(\theta_{_{\mathrm{QWP}}})\cos(\theta_{\mathrm{F}})|.
\end{equation}

\section*{Appendix B: Characterization of the B\MakeLowercase{i}$_3$F\MakeLowercase{e}$_5$O$_{12}$ sample}

\begin{figure}[h]
\includegraphics[width=0.75\columnwidth]{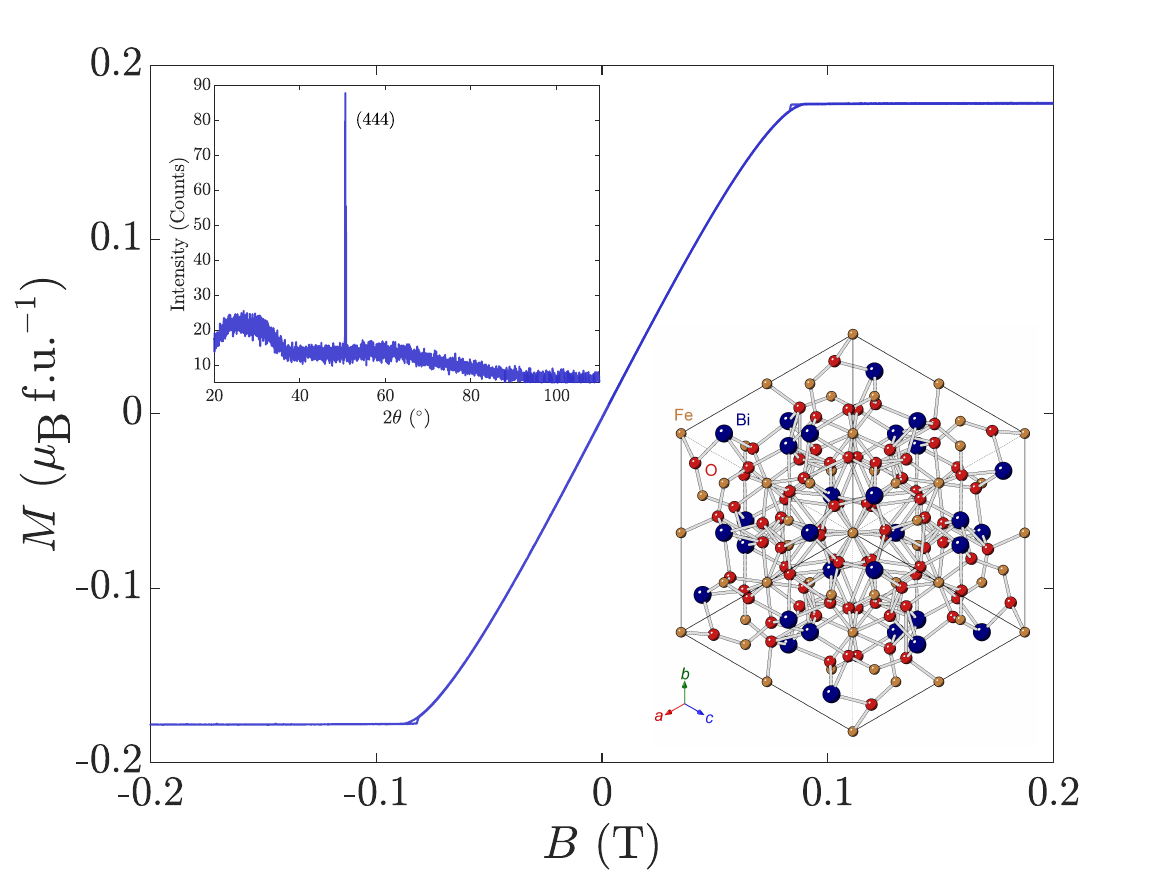}
\caption{Isothermal magnetization data as a function of applied magnetic field. Upper inset shows single-crystal $\theta$-2$\theta$ X-ray diffraction scan data and lower inset shows the cubic lattice structure of BIG.
\label{fig:MH_XRD}}
\end{figure}

We perform magnetization and X-ray diffraction (XRD) measurements for characterizing of B$_3$Fe$_5$O$_{12}$ crystal used in our measurements. Fig. \ref{fig:MH_XRD} shows the isothermal magnetization curve measured (using a superconducting quantum interference-based magnetometer by Quantum Design) as a function of externally applied magnetic field at 300 K. The data shows that magnetization saturates when the magnetic field is approximately 100 mT, which is consistent with the magnetic field dependent Faraday rotation results shown in the inset of Fig. \ref{fig:Verdet_const}(a). The upper inset of Fig. \ref{fig:MH_XRD} shows the $\theta$-2$\theta$ XRD scanning depicting only the (444) peak, that confirms the single crystallinity of the BIG sample \cite{flament2002magneto,adachi2002magnetic}. The unit cell is cubic with $a$ = 12.493 $\AA$ (see the lower inset of  Fig. \ref{fig:MH_XRD}).

\section*{Appendix C: Coherence length of the idler photons at 1540 nm}

\begin{figure}[h]
\includegraphics[width=0.65\columnwidth]{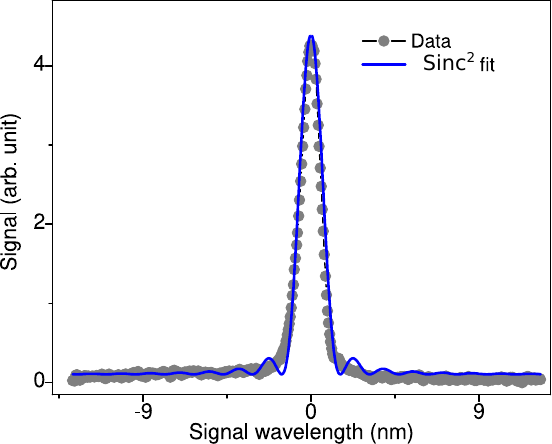}
\caption{Optical spectrum of the signal photons at 813 nm. The data is fitted to a Sinc$^2$ curve. The linewidth is 1.40$\pm$0.02 nm, which is the QPM bandwidth of the photons at near-IR range. 
\label{fig:Signal_spectrum}}
\end{figure}

The spectrum of the signal SPDC photons generated from PPLN crystal with 7.4 $\mu$m poling periodicity and  T=126$^{\circ}$C applied temperature is shown in Fig. \ref{fig:Signal_spectrum}. The central wavelength of the spectrum is 813$\pm$0.2 nm, the bandwidth is 1.40$\pm$0.02 nm, which gives the 0.48$\pm$0.01 mm coherence length.

\begin{figure}[h]
\includegraphics[width=0.75\columnwidth]{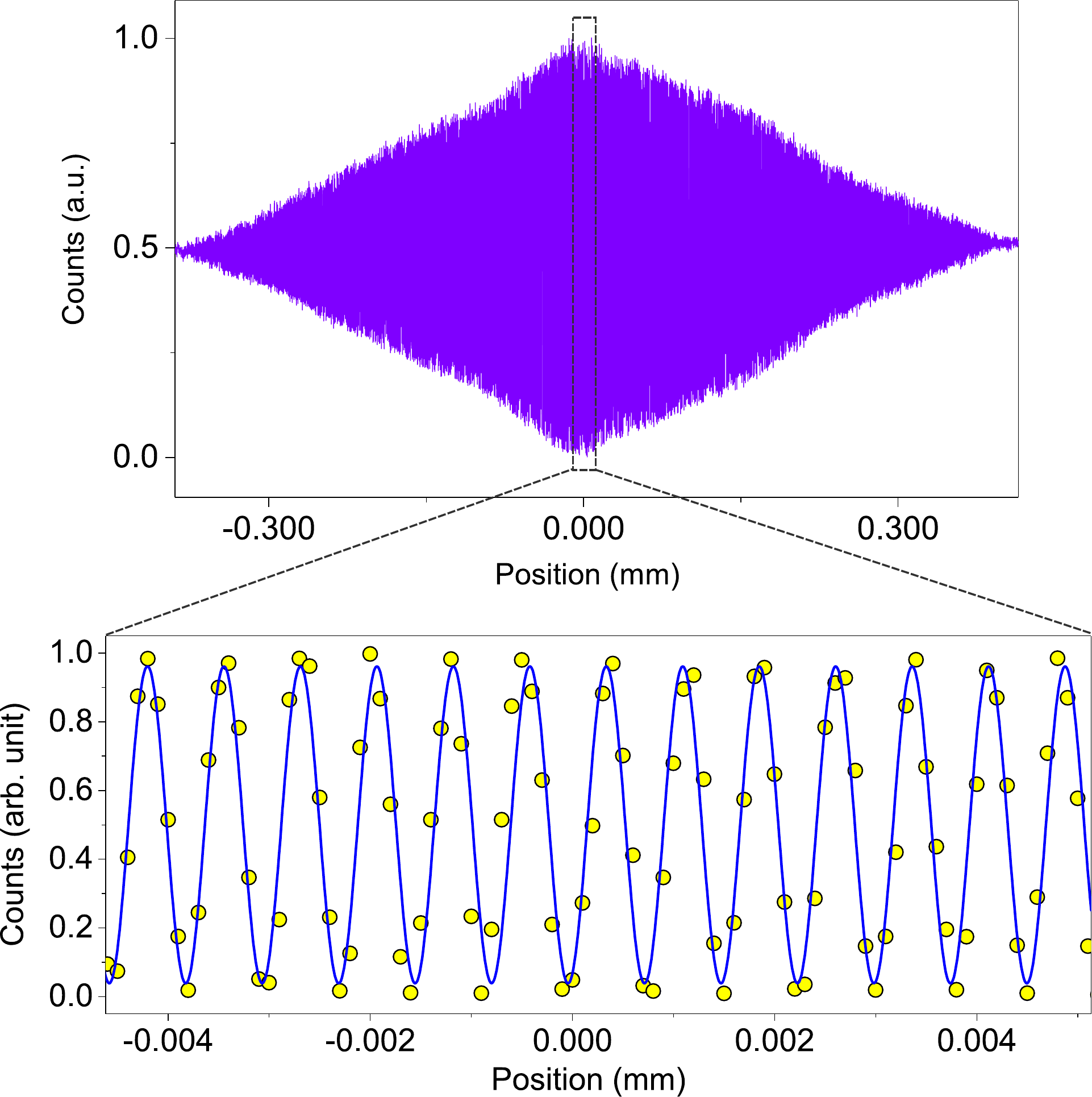}
\caption{Interference fringes depending on the additional phase introduced by translating mirror $M_i$ of the interferometer when the sample is absent. The inset at the bottom shows the zoomed in interference data around the balanced position of the interferometer and its fit to cosine function. 
\label{fig:Coh_length_no_BIG}}
\end{figure}

Intensity of the signal photons captured as a function of the translation of M$_\emph{i}$ (MTS25/M, Thorlabs Inc.) shows the interference of the signal photons in Fig. \ref{fig:Coh_length_no_BIG}. The interference is observable within the coherence length of $\approx$ 0.44 of the SPDC photons, which is twice the width of the envelope (due to double pass configuration) taken at the $1/e$ amplitude decay of the interference fringes \cite{lopez2012coherence}, see Fig. \ref{fig:Coh_length_no_BIG}.  
In the balanced position of the interferometer, the visibility of the interference is maximum. The inset at the bottom of Fig. \ref{fig:Coh_length_no_BIG} shows an enlarged view of the interference fringes close to the balanced position of the interferometer. Due to the double-pass along the arms of the Michelson-type interferometer, the periodicity of the interference fringes corresponds to half of the idler wavelength. Fitting the data to a cosine curve gives the periodicity of the interference fringes, which is 759 nm. We attribute the error in this value to the irreversibility caused by the backlash error of the transnational stage and other mechanical components.\\

\begin{figure}[h]
\includegraphics[width=0.75\columnwidth]{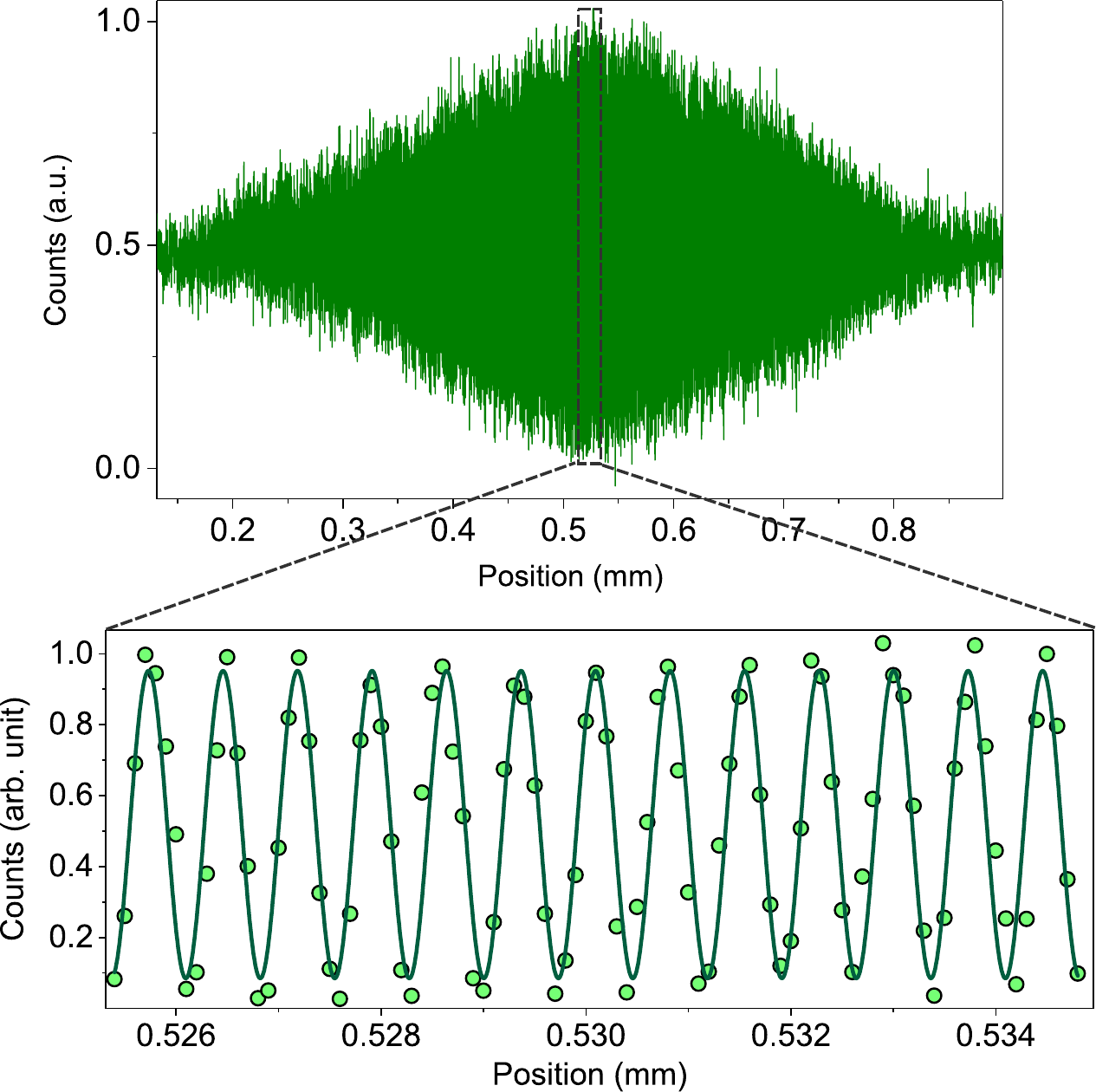}
\caption{Interference fringes depending on the translation of the mirror $M_i$ when the sample is present. The interference fringes at the optical path length difference of 0.53 mm are shown in the inset at the bottom. The fitting is performed using cosine function. 
\label{fig:Coh_length_with_BIG}}
\end{figure}

The interference fringes after the insertion of the BIG crystal into the idler arm of the interferometer are shown in Fig. \ref{fig:Coh_length_with_BIG}. Due to change in the optical path length, the balanced position of the interferometer is shifted by 0.53 mm. This change of the optical path length corresponds to the 0.38 mm thickness of the BIG sample and depends on the wavelength of the idler photons (as refractive index of a medium varies with wavelength of the light according to the Sellmeier relation \cite{ sellmeier1872ueber}). The inset at the bottom in Fig. \ref{fig:Coh_length_with_BIG} shows an enlarged view of the interference fringes and their fit to cosine curve. 

\section*{Appendix D: Visibility of the interference without the sample}

\begin{figure}[h!]
\includegraphics[width=0.65\columnwidth]{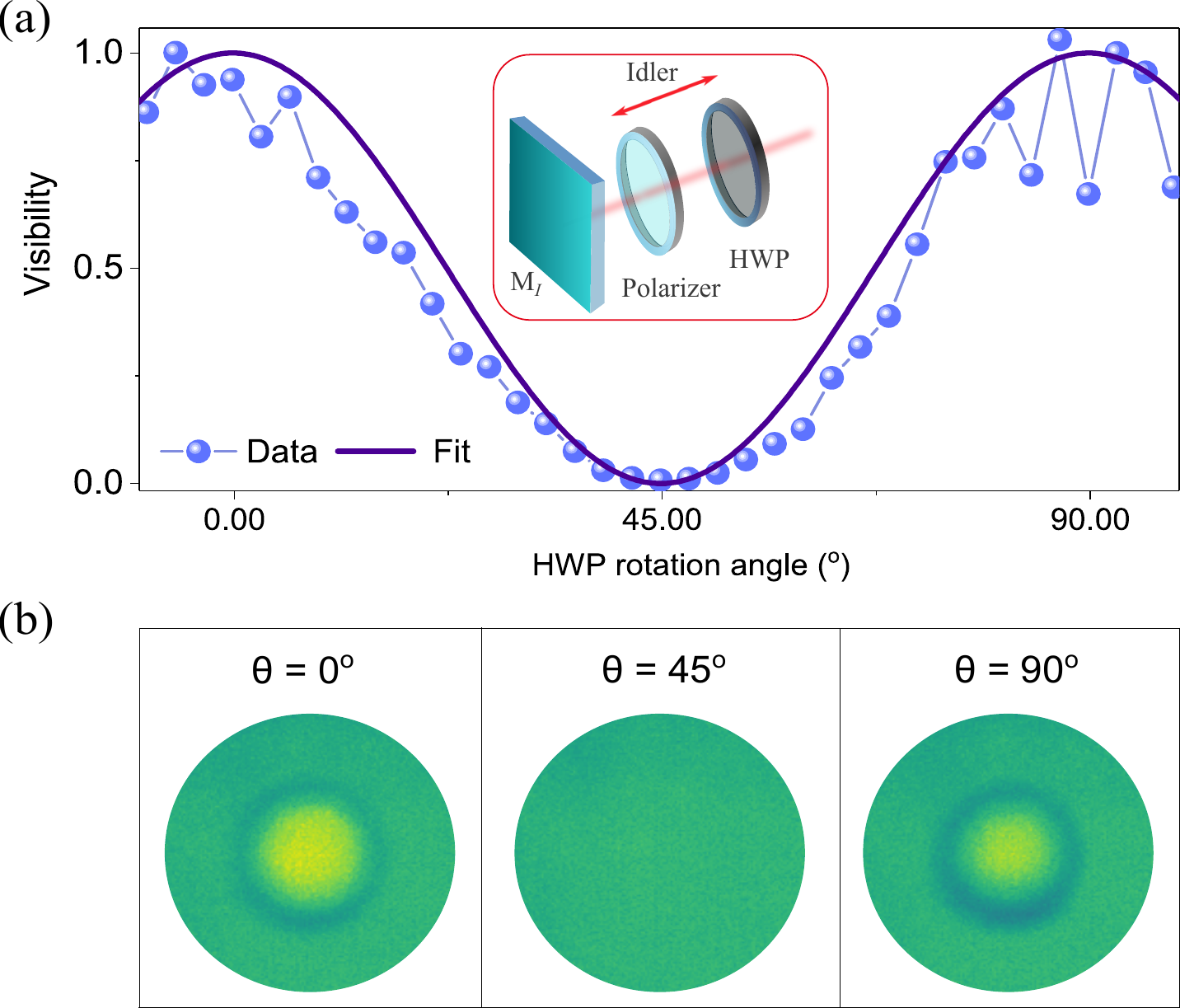}
\caption{(a) The visibility of the interferene at the absence of the sample as a function of HWP orientation. (b) The examples of the constructive interference pattern captured by sCMOS camera at few orientation angles of the HWP. 
\label{fig:V_without_BIG _HWP}}
\end{figure}

In case I, in the absence of BIG sample, the visibility of interference fringes for signal photons depends on the orientation of the HWP and polarizer (see the inset in Fig. \ref{fig:V_without_BIG _HWP}). Since the polarizer is aligned with the initial polarization of the idler photons, the visibility of the interference is given by the orientation of the HWP only. For each rotation angle of the HWP we introduce additional phase into the interferometer by translating mirror M$_\emph{i}$ over a few periods of $\lambda_{\mathrm{idler}}$, capture the interference fringes and estimate the visibilities, see Fig. \ref{fig:V_without_BIG _HWP}.  The constructive interference pattern of the signal photons captured by the sCMOS camera at angles of HWP at 0$^{\circ}$, 45$^{\circ}$ and 90$^{\circ}$ are shown in Fig. \ref{fig:V_without_BIG _HWP}(b).

\begin{figure}[h!]
\includegraphics[width=0.65\columnwidth]{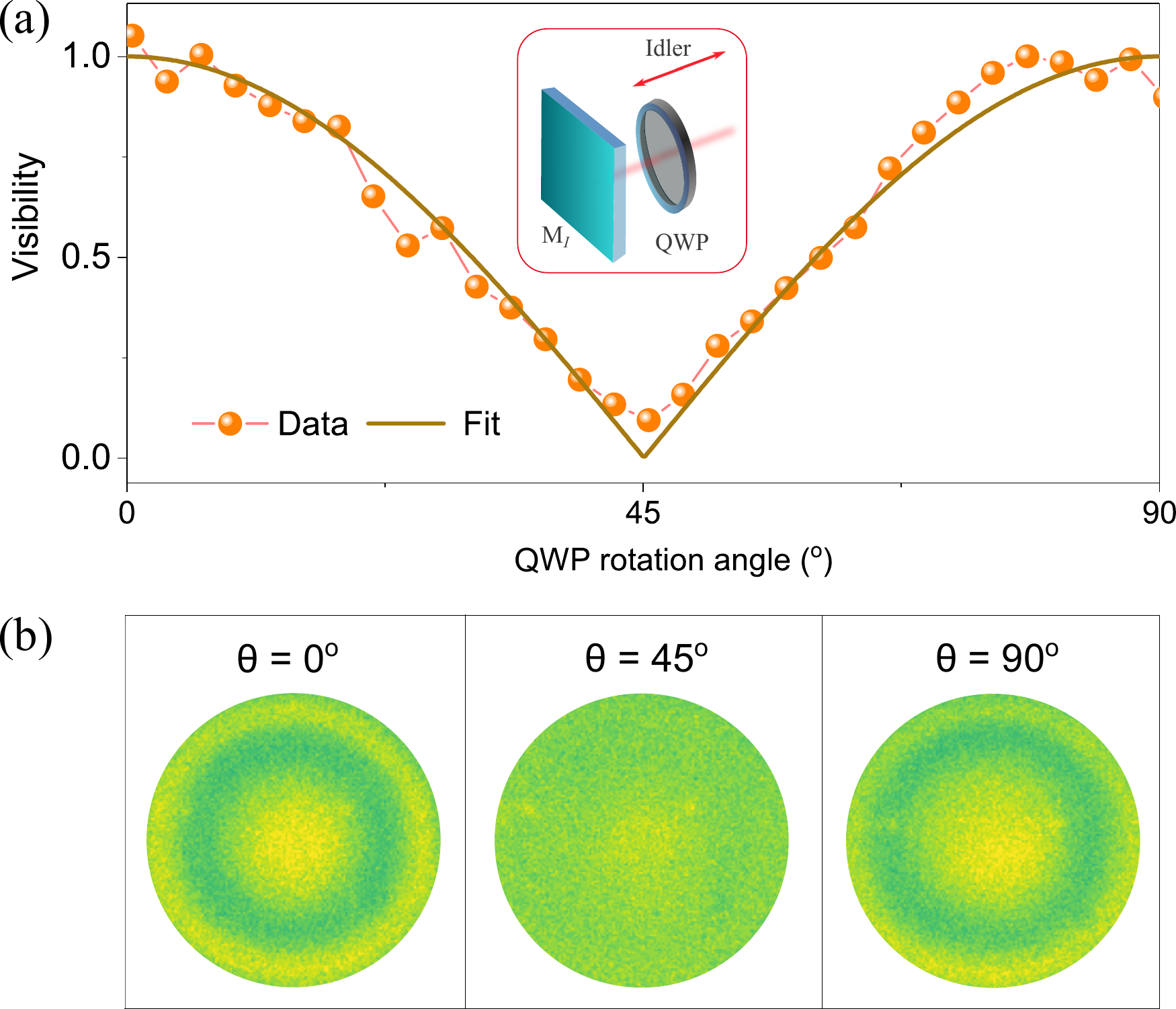}
\caption{(a) The visibility of the interference at the absence of the sample as a function of QWP orientation. (b) The examples of the constructive interference pattern captured by sCMOS camera at few orientation angles of the QWP.
\label{fig:V_without_BIG_QWP}}
\end{figure}

In case II, when we replace the HWP by a QWP and remove the polarizer (see the inset), the balance condition for the interferometer changes. We balance the interferometer by translating the mirror M$_\emph{i}$. Next, we record the change in visibility of the interference depending on the orientation of QWP in the absence of the BIG sample (see Fig. \ref{fig:V_without_BIG_QWP}). Likewise, the constructive interference pattern of the signal photons at orientation angles of QWP at 0$^{\circ}$, 45$^{\circ}$ and 90$^{\circ}$ are shown  Fig. \ref{fig:V_without_BIG_QWP}(b). Both of the obtained visibility data is in a good agreement with the theoretical analysis given by Eq. \ref{eq:Vis_HWP} (case I) and Eq. \ref{eq:M_QWP} (case II).

\section*{Appendix E: Faraday rotation angles measured via nonlinear interferometry}

\begin{table}[h!]
\begin{ruledtabular}
\caption{Applied magnetic field strengths and measured Faraday rotations.}

\begin{tabular}{|c|c|}

\thead{Magnetic field (case I)} & \thead{Faraday rotation(case I)}\tabularnewline
\hline
\hline

$75\pm8$ mT & $34.08^{\circ}\pm0.22^{\circ}$ \tabularnewline

$141\pm15$ mT & $44.34^{\circ}\pm0.46^{\circ}$ \tabularnewline

\hline
\hline
\thead{Magnetic field (case II)} & \thead{Faraday rotation(case II)}\tabularnewline
\hline
\hline

$36.5\pm3.9$ mT & $16.46^{\circ}\pm0.47^{\circ}$ \tabularnewline

$75\pm8$ mT & $31.24^{\circ}\pm0.15^{\circ}$ \tabularnewline

$106\pm4.1$ mT & $42.31^{\circ}\pm0.08^{\circ}$ \tabularnewline

$141\pm15$ mT & $44.9^{\circ}\pm0.45^{\circ}$ \tabularnewline

\end{tabular}

\label{table:QWP_BIG_rot}
\end{ruledtabular}
\end{table}

Here in table \ref{table:QWP_BIG_rot} we show the experimentally measured Faraday rotations depending on the applied magnetic fields in case I and II.

\section*{Appendix F: Wavelength dependent Faraday rotation}

To perform metrology of the Verdet constant and the Faraday rotation at the saturation magnetic field over a broad IR wavelength range, we generate the idler photons in the range of 1540 nm to 2141 nm by pumping the PPLN through different quasi phase-matched channels and setting the corresponding temperatures. We measure the signal (visible/near-IR) photon spectra using a spectrometer (HR4000, Ocean Optics) and determine the signal wavelengths. Following the energy conservation, we determine the wavelengths of idler photons. Table \ref{table:QPM_T_S_I} summarizes signal$\lambda_{\mathrm{idler}}$ and idler wavelengths $\lambda_{\mathrm{idler}}$, with corresponding values of the polling periods and PPLN temperatures.

\begin{table}[h]
\caption{Poling periods of the PPLN and temperatures for generating signal and idler photons at different wavelengths. The wavelength of the pump laser is 532 nm in all these cases.\\}

\begin{tabular}{|c|c|c|c|}
\hline
\thead{Polling \\ period} & Temperature & \thead{Signal wavelength \\ (detected)} & \thead{Idler wavelength\\ (Probed)}\tabularnewline
\hline 
\hline

7.4 $\mu$m & 378 K & 820 nm & 1516 nm \tabularnewline
\hline 

7.4 $\mu$m & 399 K & 813 nm & 1540 nm \tabularnewline
\hline 

7.4 $\mu$m & 423 K & 805 nm & 1568 nm \tabularnewline
\hline 

7.71 $\mu$m & 303 K & 793 nm & 1617 nm \tabularnewline
\hline 

8.36 $\mu$m & 352 K & 725 nm & 2000 nm \tabularnewline
\hline

8.36 $\mu$m & 571 K & 708 nm & 2141 nm \tabularnewline
\hline

\hline 
\end{tabular}
\label{table:QPM_T_S_I}

\end{table}

\begin{figure*}[t!]
\includegraphics[width=0.75\textwidth]{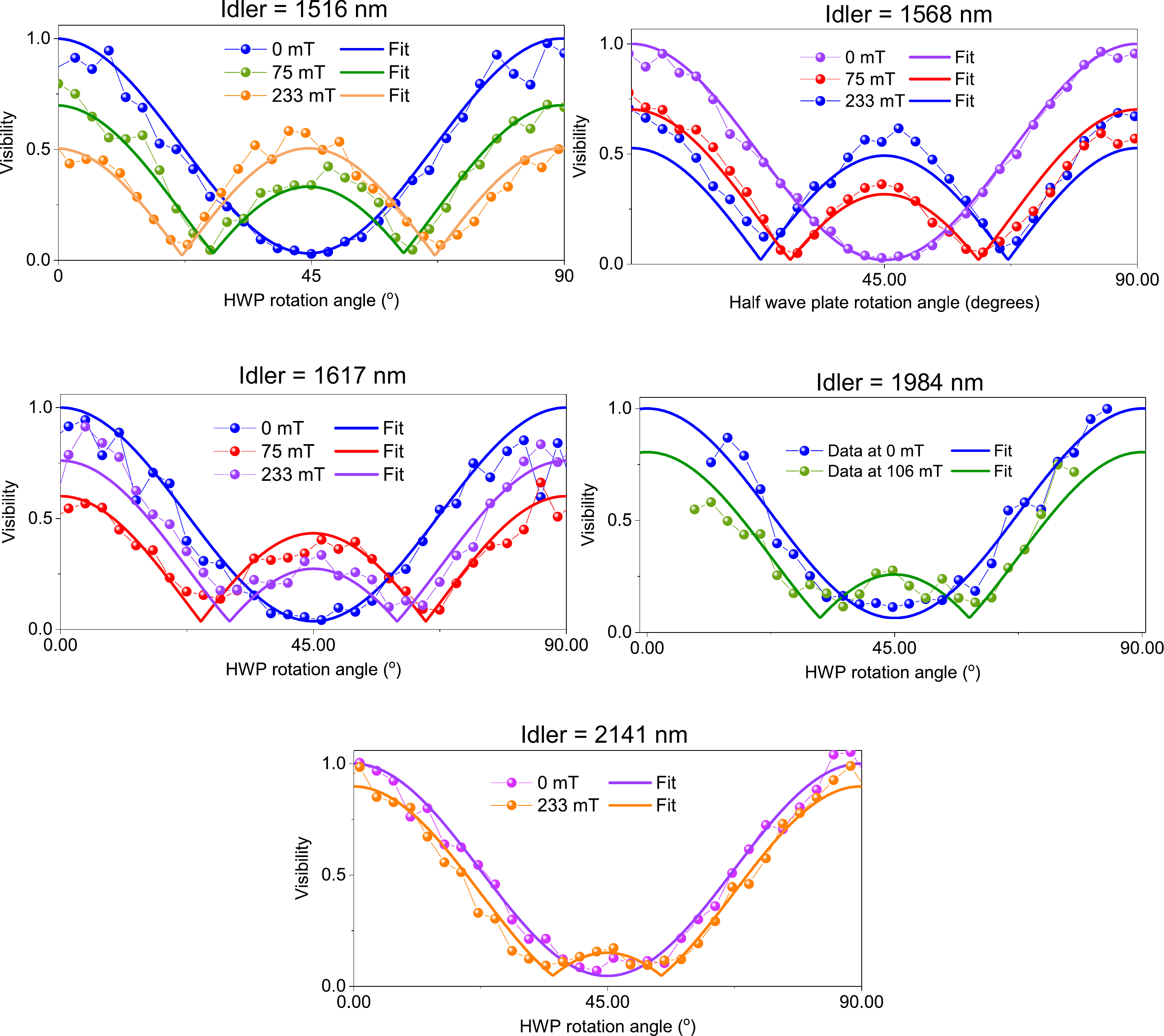}
\caption{Variation of visibilities of interference of the signal photons with the rotation angle of the HWP for different idler wavelengths. 
\label{fig:V_plots_various_WL}}
\end{figure*}

For each of these cases, we measure visibility of the interfering signal photons as a function of the input polarization of the idler photons. Figure \ref{fig:V_plots_various_WL} shows the visibilities of the interference pattern depending on HWP rotation angle (protocol for the case I) and their fit using Eq. \ref{eq:Vis_HWP}. We determine $\theta_{\mathrm{F}}$ from the fitting and estimate the Verdet constant. For the estimation of the Verdet constant we choose the magnetic fields below the saturation field. At the saturation magnetic field we calculate the Faraday rotation angles as a function of wavelength.

\bibliographystyle{apsrev4-2}
\bibliography{references}

\end{document}